\documentclass[sigconf,natbib=true,anonymous=false]{acmart}
\usepackage{amsmath,nccmath,tabularx} 
\usepackage{algorithm}
\usepackage{algorithmic}

\usepackage{color}
\usepackage{graphicx}
\graphicspath{{images/}}
\usepackage{wrapfig,lipsum}
\usepackage{tabularx}
\usepackage{graphicx}
\usepackage{lscape}
\usepackage{subcaption}
\usepackage{latexsym}
\usepackage{pifont}
\usepackage{multirow}

\newcommand{\katya}[1]{\textcolor{black}{#1}}

\usepackage{enumitem}
\setlist{leftmargin=2mm}
\usepackage[T1]{fontenc}
\usepackage[utf8]{inputenc}
\usepackage[font=small,labelfont=bf]{caption}

\usepackage{listings}
\usepackage{balance}

\usepackage{xcolor}

\colorlet{punct}{red!60!black}
\definecolor{background}{HTML}{EEEEEE}
\definecolor{delim}{RGB}{20,105,176}
\colorlet{numb}{magenta!60!black}
\lstdefinelanguage{json}{
	basicstyle=\normalfont\ttfamily,
	numberstyle=\scriptsize,
	stepnumber=1,
	showstringspaces=false,
	literate=
	*{0}{{{\color{numb}0}}}{1}
	{1}{{{\color{numb}1}}}{1}
	{2}{{{\color{numb}2}}}{1}
	{3}{{{\color{numb}3}}}{1}
	{4}{{{\color{numb}4}}}{1}
	{5}{{{\color{numb}5}}}{1}
	{6}{{{\color{numb}6}}}{1}
	{7}{{{\color{numb}7}}}{1}
	{8}{{{\color{numb}8}}}{1}
	{9}{{{\color{numb}9}}}{1}
	{:}{{{\color{punct}{:}}}}{1}
	{,}{{{\color{punct}{,}}}}{1}
	{\{}{{{\color{delim}{\{}}}}{1}
	{\}}{{{\color{delim}{\}}}}}{1}
	{[}{{{\color{delim}{[}}}}{1}
	{]}{{{\color{delim}{]}}}}{1},
}

\AtBeginDocument{%
  \providecommand\BibTeX{{%
    \normalfont B\kern-0.5em{\scshape i\kern-0.25em b}\kern-0.8em\TeX}}}

\copyrightyear{2024}
\acmYear{2024}
\setcopyright{acmlicensed}\acmConference[SIGIR '24]{Proceedings of the 47th International ACM SIGIR Conference on Research and Development in Information Retrieval}{July 14--18, 2024}{Washington, DC, USA}
\acmBooktitle{Proceedings of the 47th International ACM SIGIR Conference on Research and Development in Information Retrieval (SIGIR '24), July 14--18, 2024, Washington, DC, USA}
\acmDOI{10.1145/3626772.3657674}
\acmISBN{979-8-4007-0431-4/24/07}

\begin{document}

\title{Embark on DenseQuest: A System for Selecting the Best Dense Retriever for a Custom Collection}





\author{Ekaterina Khramtsova}
\affiliation{%
	\institution{The University of Queensland}
	\city{Brisbane}
	\country{Australia}}
\email{e.khramtsova@uq.edu.au}

\author{Teerapong Leelanupab}
\affiliation{%
	\institution{The University of Queensland}
	\city{Brisbane}
	\country{Australia}}
\email{t.leelanupab@uq.edu.au}

\author{Shengyao Zhuang}
\affiliation{%
	\institution{CSIRO}
	\city{Brisbane}
	\country{Australia}}
\email{shengyao.zhuang@csiro.au}

\author{Mahsa Baktashmotlagh}
\affiliation{%
	\institution{The University of Queensland}
	\streetaddress{4072 St Lucia}
	\city{Brisbane}
	\country{Australia}}
\email{m.baktashmotlagh@uq.edu.au}

\author{Guido Zuccon}
\affiliation{%
	\institution{The University of Queensland}
	\streetaddress{4072 St Lucia}
	\city{Brisbane}
	\country{Australia}}
\email{g.zuccon@uq.edu.au}

\begin{abstract}
In this demo we present a web-based application for selecting an effective pre-trained dense retriever to use on a private collection. Our system, \texttt{DenseQuest}, provides unsupervised selection and ranking capabilities to predict the best dense retriever among a pool of available dense retrievers, tailored to an uploaded target collection.
\texttt{DenseQuest} implements a number of existing approaches, including a recent, highly effective method powered by Large Language Models (LLMs), which requires neither queries nor relevance judgments.  The system is designed to be intuitive and easy to use for those information retrieval engineers and researchers who need to identify a general-purpose dense retrieval model to encode or search a new private target collection. Our demonstration illustrates conceptual architecture and the different use case scenarios of the system implemented on the cloud, enabling universal access and use. \texttt{DenseQuest} is available at \url{https://densequest.ielab.io}.
\end{abstract}


\maketitle

\vspace{0.2cm}
\section{Introduction}
\vspace{0.1cm}




Recent advances in language modelling for Information Retrieval (IR) have resulted in the development of numerous  dense retrievers (DRs) with diverse architectures, sizes, training mechanisms and scoring functions~\cite{tonellotto2022lecture,10.1145/3637870}. Selecting the most appropriate dense retriever for a specific collection can yield significant performance improvements compared to selecting a sub-optimal one. However, in many real-world scenarios, direct performance evaluation of these models is impractical due to the lack of available relevance judgments (or labels). Indeed, acquiring judgments for private collections is not only time-consuming and costly, but often also requires domain-specific knowledge and expertise. This raises a question: How can one effectively choose the best dense retriever to deploy on their collection when no labeled data is available?

Currently, search engine practitioners assessing the capabilities of DRs often refer to published leaderboards, which can be categorized into in-domain and zero-shot. In-domain leaderboards assess models that are trained and evaluated on different subsets of the same collection, such as leaderboards for HotpotQA~\cite{yang2018hotpotqa} and MS MARCO~\cite{bajaj2016ms}. In contrast, zero-shot leaderboards, such as  MTEB~\cite{muennighoff-etal-2023-mteb} and BEIR~\cite{thakur2021}, evaluate the performance of models trained on one task but tested on data from a different collection, domain or task.  However, neither of these systems provides recommendations for new, private data; they only report the performance on existing publicly available collections. To the best of our knowledge, our application is the first to offer tailored recommendations for individual private collections without requiring the prior acquisition of relevance judgments.

Bridging this gap, our proposed web-based system, \texttt{DenseQuest}, employs a diverse range of unsupervised performance evaluation methods (which we refer to as DR selection methods). These methods can be categorized as either per-query (query performance prediction, or QPP~\cite{Faggioli2023QPP}) or per-collection~\cite{khramtsova2023}. QPP approaches aim to predict the performance of each individual query within a model, whereas per-collection methods seek to predict the average performance across all queries, focusing on model comparison rather than query comparison. We incorporate methods from both categories into \texttt{DenseQuest}, including Score-based  (\texttt{SMV}~\cite{Tao2014SMV}, \texttt{NQC}~\cite{Shtok2012NQC}, \texttt{$\sigma$}~\cite{Cummins2011SigmaMax})  and Fusion-based~\cite{Shtok2016QueryPP} QPP from the first category, and MSMARCO-based ranking~\cite{bajaj2016ms},  MTEB-based ranking~\cite{muennighoff-etal-2023-mteb}, Binary Entropy~\cite{khramtsova2023}, Query Alteration~\cite{khramtsova2023}, and LARMOR~\cite{khramtsova2024larmor} from the second category.



\texttt{DenseQuest} makes these unsupervised performance evaluation methods accessible and easy to deploy.
The system is designed for search engine practitioners and researchers who need to make an informed decision on the selection of dense retrievers, based exclusively on an unjudged private collection. \texttt{DenseQuest} consolidates existing state-of-the-art methods for DR selection and ranking. It enables users to upload their collections, choose their preferred model selection algorithm, and identify the best DR suited to their specific collection. 
Additionally, users have the option to download the trained checkpoint for the selected model, accompanied by 
instructions for its use, thereby simplifying the deployment process.

\section{DenseQuest}


\vspace{-0.1cm}
\subsection{Architecture and Workflow}
The architecture of \texttt{DenseQuest} consists of two individual Docker containers (see Figure~\ref{fig:arch}) separately deployed in cloud instances. These include $(i)$ a web-based front-end interface built with Vue.js and Tailwind CSS and $(ii)$ a back-end for REST APIs built with the Python web framework Django. In the front-end container, we use Nginx as a web server to serve static content. In the back-end container, we also install an SQLite database for storing information on user application usage, uploaded collections, processed jobs, ranking results, etc. Besides, the back-end part comprises two principal components, implemented in Python, including the task queue management and the core of \texttt{DenseQuest}.

The workflow of our system is as follows. Once a user uploads a collection (step 1), a job is created as a request to find the optimal DR model for a given collection. The front-end then passes his request to the back-end (step 2). The Rest API receives the request, stores information about the job and collection in the SQLite database, and adds the processing job to a queue (step 3). Once the computational resources have been allocated to that job, the \texttt{DenseQuest} core executes consecutively the collection encoding and the model selection (step 4). Upon completion, the \texttt{DenseQuest} core stores the results for model selection and returns them to the RestAPI (step 5), which in turn updates the status of the job and returns it to the website for visualisation (step 6). The user then has the option to download the model checkpoint (step 7).


\vspace{-0.1cm}
\subsection{Infrastructure}

In terms of cloud infrastructure, we use Amazon Web Services (AWS) to deploy, manage, and power our  \texttt{DenseQuest} application. For the front-end that serves only a web server, we employ a EC2 general purpose instance, with a series of M7i (m7i.large: 2 vCPUs and 8 GB instance memory.) The back-end, which requires GPU computation, utilises a EC2 accelerated computing instance, with a currently selected series of G4\footnote{For the purpose of this demo, we 
select a cost-effective instance type; a more powerful one should be chosen to scale to higher computing demands.
} (g4dn.xlarge: 1 GPU of T4 Tensor,  16 GB GPU memory, 4 vCPUs, and 16 GB instance memory.) This specification can be simply scaled up without further modification of application environments to other compatible series with higher hardware accelerators and their number of units, such as G5 or P4, which come with NVIDIA A10G and A100 GPUs, respectively. The AWS S3 scalable storage is used to store user-uploaded collections, data embedding, dense retrievers, etc.

\vspace{-0.1cm}
\subsection{Task Queue Management}

For managing the task queue, we adopt Celery Worker for asynchronous task execution and RabbitMQ as a message broker. 

There are two primary reasons for introducing a task queue management.
First, when HTTP requests take a significant amount of time, 
message queues are necessary to handle and respond to multiple requests concurrently coming before timeout. Second, we apply these tools to distribute and balance the workload of tasks from user requests across workers using queues specifically designated for tasks with GPU or CPU usage. 
\katya{One queue is allocated for computationally intensive GPU-based tasks, whereas the other focuses on quicker CPU-based tasks.}
We can flexibly configure the number of workers and queues according to the resources (e.g., the number of vCPUs and GPUs) available during implementation.

\begin{figure}[t!]
	\includegraphics[width=0.46\textwidth]{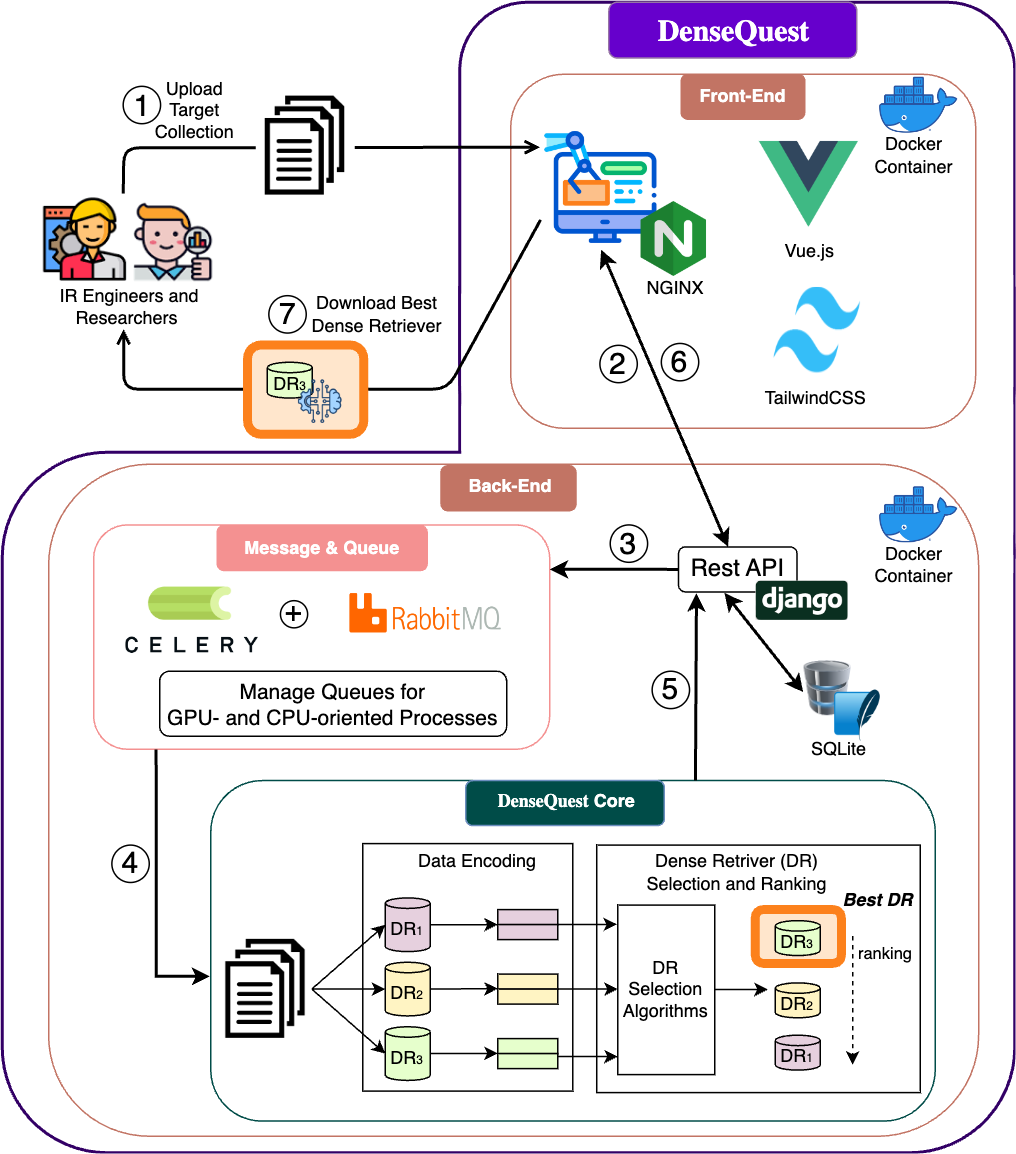}
	\vspace{-0.25cm}
	\caption{ Overview of the architectural components of \texttt{DenseQuest}}
	\label{fig:arch}
	\vspace{-0.5cm}
\end{figure}

\begin{figure*}
		\includegraphics[width=0.77\textwidth]{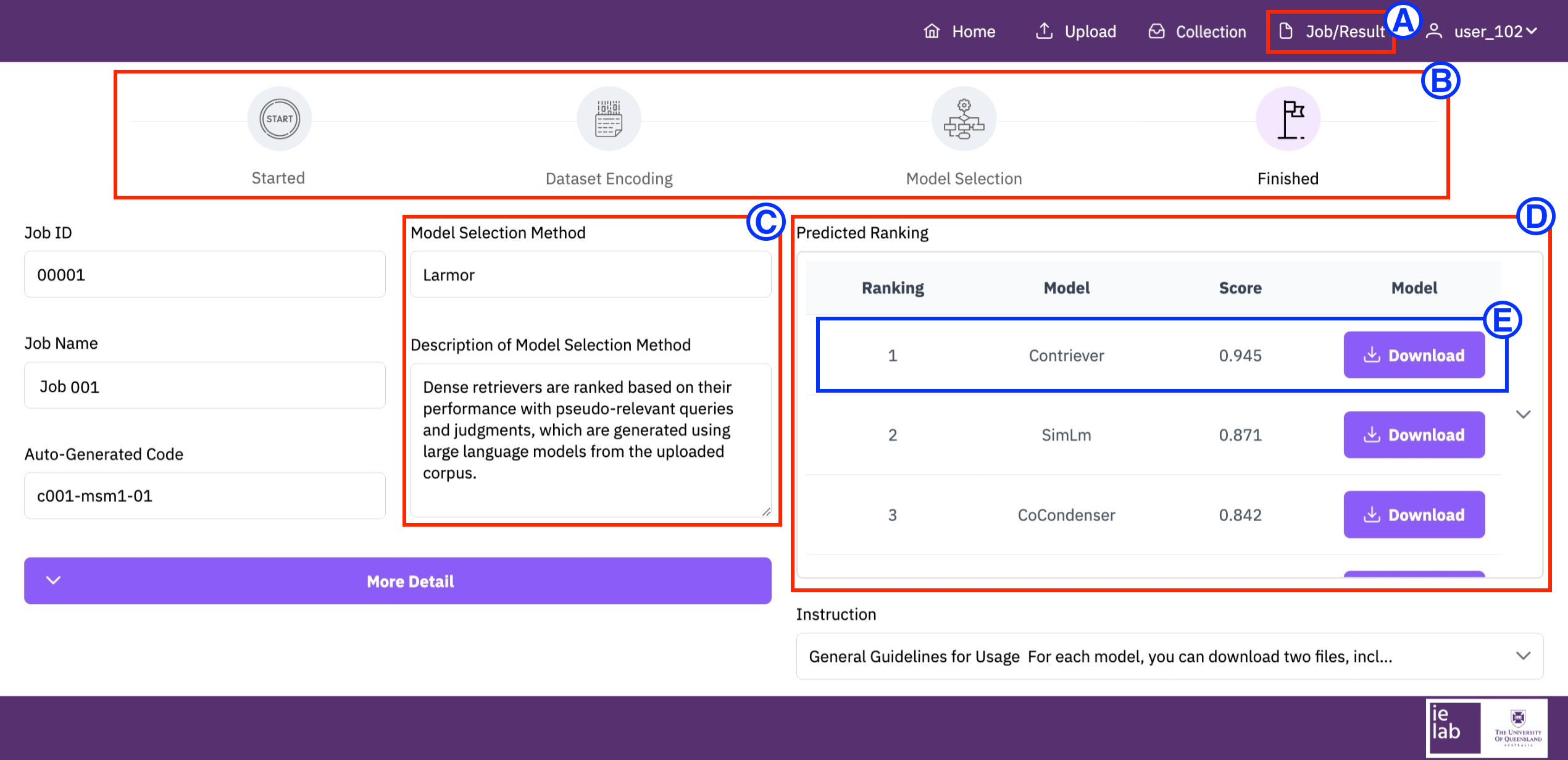}
		\vspace{-0.25cm}
		\caption{ \texttt{DenseQuest} result page for a target collection uploaded by a user. Note that for publication purposes, the layout of the UI presented here has been modified and differs slightly from that of an actual UI, while retaining all functionalities. }
		\label{fig:ui}
		\vspace{-0.276cm}
	\end{figure*}

\vspace{-0.1cm}
\subsection{The Core of DenseQuest}

The core functionality of \texttt{DenseQuest} can be divided in two major components: Data encoding and DR selection and ranking.

The first component involves forwarding the corpus through all the DRs from the model pool and storing model-specific embeddings.  This procedure requires a GPU and
represents the most time-consuming step of the pipeline. For efficiency, we perform this operation only once and save the embeddings for each user. 

The second component involves ranking the models, where each DR selection method assigns a single score to each DR, which will be further used for ranking. We consider the following methods:

\begin{enumerate}
	\item   Binary Entropy~\cite{khramtsova2023} assesses model uncertainty using the binary entropy, calculated for each query based on the respective scores.
	\item Query Alteration~\cite{khramtsova2023} evaluates the model's sensitivity to masking query terms by measuring the variability in the scores of the retrieved documents. 
	\item Score-based QPP methods \texttt{SMV}~\cite{Tao2014SMV}, \texttt{NQC}~\cite{Shtok2012NQC}, \texttt{$\sigma$}~\cite{Cummins2011SigmaMax}. 
	Every query is assigned a value calculated from the standard deviation of scores of the corresponding retrieved documents. The model is then characterized by the average value across all queries.
	\item Fusion-based QPP ~\cite{Shtok2016QueryPP}. First, a pseudo-relevant ranking is created by fusing the rankings from all the models in the pool.  Then each model is assigned a value based on the similarity of its produced ranking to the pseudo-relevant ranking.
	\item   MSMARCO- ~\cite{bajaj2016ms} and  MTEB-based ranking~\cite{muennighoff-etal-2023-mteb}: models are ranked based on their position on MSMARCO and MTEB leaderboards.
	\item LARMOR~\cite{khramtsova2024larmor}: evaluates each model based on pseudo-relevant queries and pseudo-relevant judgments, which are generated using Large Language Models (LLMs).
\end{enumerate}

Note that neither method requires relevance judgments, making them unsupervised model selection methods. Furthermore, methods from the last two groups (5 and 6) are query-free. Additionally, only LARMOR and Query Alteration require GPU; all other methods can be computed rapidly using only CPUs. 

%
%
%
%
%

\vspace{-0.1cm}
\section{Demonstration}
\katya{ 
In this section, we introduce the \texttt{DenseQuest} web application, 
accessible through a web interface at  \url{https://densequest.ielab.io}.}

\begin{figure}
	\includegraphics[width=0.43\textwidth]{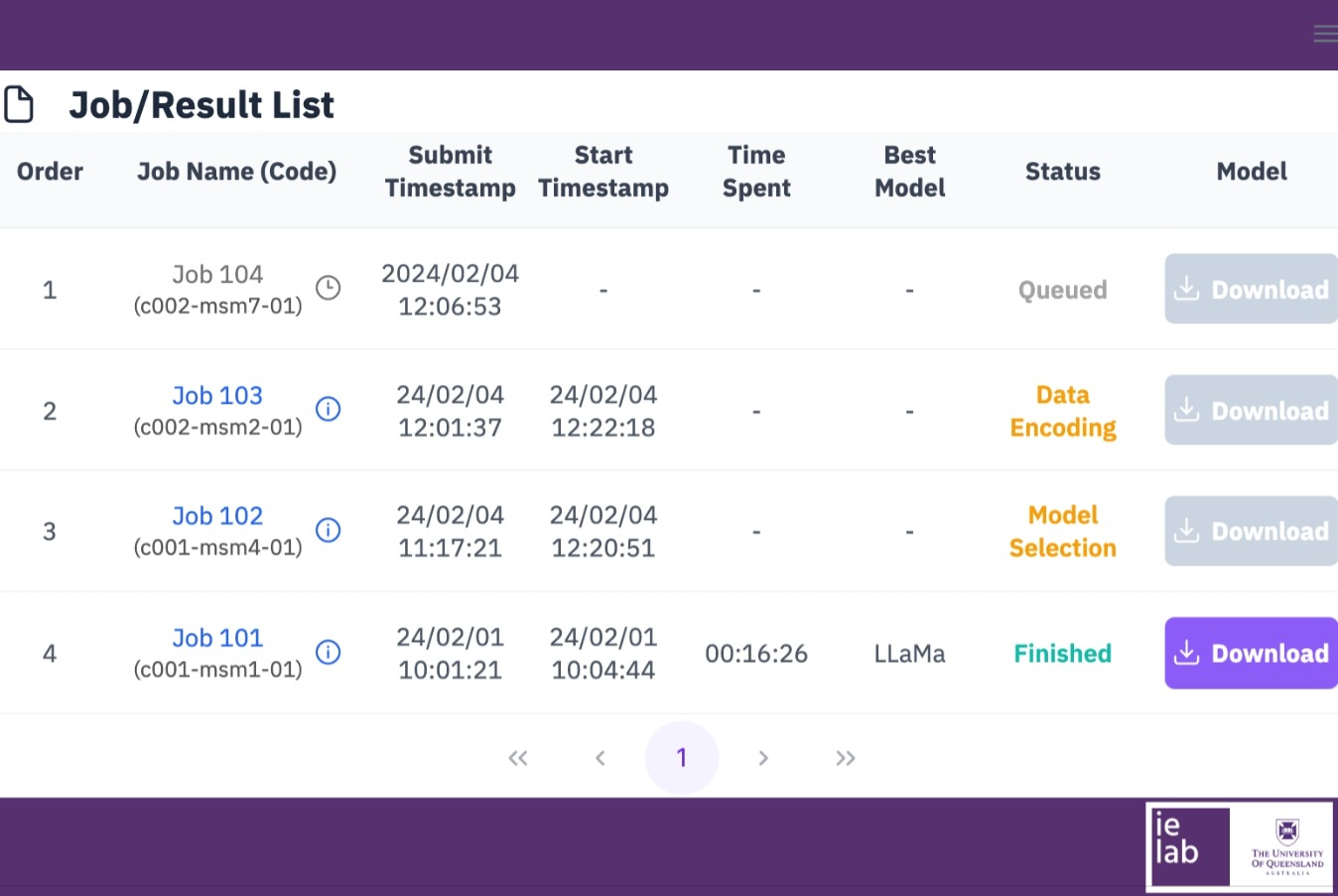}
	\vspace{-0.25cm}
	\caption{ \texttt{DenseQuest} job/result list, as uploaded by a user. }
	\label{fig:job_status}
	\vspace{-0.6cm}
    
\end{figure}

    \katya{ 
    IR engineers and researchers interested in using \texttt{DenseQuest} must first register to create user accounts.} They can then upload their custom collection to find \katya{the most suitable DR for their needs (step 1 in Figure~\ref{fig:arch}.) For each corpus, users have the flexibility to experiment with various model selection methods provided by the system. }
Each submission is considered to be a job, comprising one or more tasks to be processed at the back-end. Our system supports having multiple jobs for each user. The status of each job and the associated timestamps are displayed in the table, as seen in Figure~\ref{fig:job_status}.

In addition, the result page for each job can be accessed under the menu of Job/Result $(A)$ of Figure~\ref{fig:ui} by clicking a job name from Figure~\ref{fig:job_status}, such as ``Job 101''. This page summarizes the selection and ranking results, produced by \texttt{DenseQuest}. 
In this interface, the progress bar $(B)$ highlights the current status of the job, after being popped off a queue from ``Started'' to ``Finished.'' The progress bar depicts two major interval steps, including ``Dataset Encoding'' and ``Model Selection.'' 
\katya{Additionally, the interface provides a detailed description of the chosen model selection method in section (C).}
Upon job completion, a table $(D)$ lists dense retrievers ranked in decreasing order of suitability to a given collection. 
Users can finally download any dense retriever model, particularly the best one $(E)$ as suggested by \texttt{DenseQuest} (step 7 in Figure~\ref{fig:arch}.) 
The downloadable zip file contains the trained model checkpoint and code for facilitating model deployment, with each model's code featuring a unique class that loads the model and implements the functions \texttt{encode\_queries} and \texttt{encode\_corpus}. Below, we present the parent class for all custom models.

\vspace{-0.33cm}
\begin{lstlisting}[language=python, label={code:avg}]

<@\textcolor{blue}{class}@> CustomDEModel:
    <@\textcolor{blue}{def}@> __init__(self, **kwargs):
        self.query_encoder, self.doc_encoder = <@\textcolor{blue}{None}@>, <@\textcolor{blue}{None}@>
        self.score_function = <@\textcolor{blue}{None}@>
    <@\textcolor{blue}{def}@> encode_queries(self, queries: List[str],
                batch_size: int, **kwargs) -> np.ndarray: <@\textcolor{blue}{pass}@>        
    <@\textcolor{blue}{def}@> encode_corpus(self, corpus: List[Dict[str, str]],
                batch_size: int, **kwargs) -> np.ndarray: <@\textcolor{blue}{pass}@>                 
\end{lstlisting}

%

%

The current pool of DR models supported by \texttt{DenseQuest} includes those utilized in LARMOR~\cite{khramtsova2024larmor}, which do not require data-specific instructions.

\vspace{-0.1cm}
\section{Utility and Usage/Implication}

%

\texttt{DenseQuest} is a user-friendly and accessible system that does not require prior expertise in dense retrievers or performance prediction, making it beneficial for search engine practitioners, engineers, researchers, and a broad spectrum of users.  The only requirement is to format the data into the standard BEIR-compatible format.

\begin{table*}[t]
	\caption{Time Efficiency Comparison. The values represent the time required for each method to rank all models for the NFCorpus~\cite{boteva2016} collection from the BEIR~\cite{thakur2021} benchmark. All computations are performed using one H100 GPU. 
 \vspace{-10pt}
 }
		\centering
  {\renewcommand{\arraystretch}{1.2}%
		\begin{tabular}{c|c|c|c|c|c|c|c|c|c}
			\hline
			&  Binary Entropy & Query Alteration & \texttt{SMV} & \texttt{NQC} & $\sigma$ & Fusion & MSMARCO & MTEB & LARMOR \\
			\hline
                Time (min) & 3.15 & 49.62 & 0.22 & 0.22 & 11.88 & 0.95 & 0.01 & 0.01 & 257
                \\ \hline   
		\end{tabular}
  }
	\label{tab:time}
	\vspace{-8pt}
\end{table*}

Our system is adaptable, catering to diverse user requirements by offering a variety of model selection methods. If users have only the corpus, they can choose from methods that do not depend on queries, such as LARMOR, MSMARCO, or MTEB based ranking. On the other hand, access to both corpus and queries allows for experimentation with all available methods.

Each model selection method emphasizes different aspects of dense retriever performance. For example, the query alteration approach reflects the robustness of DRs to input perturbations; the fusion method favours the most consistent model with respect to fused pseudo-relevant ranking; while LARMOR assesses models using artificially generated queries and relevance judgments, and therefore its performance is conditional on the generation quality.
This diversity enables users to choose the most fitting method according to the particular aspect of model robustness they prioritize.

In terms of predicting performance, Fusion, LARMOR, and MTEB-based ranking stand out from previous work as particularly effective in identifying the best DR with respect to its nDCG@10~\cite{khramtsova2023,khramtsova2024larmor}. 
When analyzing the BEIR collection, the Fusion approach can identify the optimal model for the Trec-News collection, while MTEB-based ranking excels on NFCorpus and DBPedia-Entity, and LARMOR is most effective for Quora. 

On the other hand,  if the user's task significantly deviates from the training task (e.g., argument retrieval) or there is a large conceptual domain shift (e.g., collections featuring specialized terminology from the medical or legal fields), Fusion and LARMOR may not identify the most specialized model.  
For example, for TREC-COVID, Fusion selects a model with a 20.09\% drop in nDCG@10 from the best DR. In such cases, methods focusing on individual model score distributions, like the QPP-based WIG method and Binary Entropy, may be preferable. 
The computational times for each method are presented in Table \ref{tab:time}\footnote{For LARMOR, we used FlanT5-large for generating queries and judgments.}; for the effectiveness comparison refer to \cite{khramtsova2024larmor}.

In practice, there is often an overlap among the top-predicted models across different methods. Therefore, in order to make an informed choice, we suggest experimenting with multiple model selection approaches and deploying top-predicted models recommended by these methods.

\vspace{-0.15cm}
\section{Future Functions}
\label{future}

%
We plan to introduce several new functionalities into \texttt{DenseQuest}.

The first enhancement involves enabling users to upload their own model selection methods. Integrating this feature into our existing infrastructure will not require significant modifications.  All current methods follow the same structure: they assign a single score to each DR,  and the DRs are then ranked in either ascending or descending order based on this score.
Therefore, any new model selection approach adhering to this format can be incorporated into our system. This functionality would support researchers in comparing their new method against existing DR selection methods, thus accelerating research advances.

The second functionality we seek to develop in future would allow users to submit their own DRs for evaluation. This however presents a considerable challenge in terms of system adaptation. The state-of-the-art DRs display a wide variety of architectures and inference procedures. To facilitate integration, we have standardized the encoding and scoring across all supported models. However, integrating new models is not straightforward, as some require distinct procedures for query and corpus embedding (e.g., custom task-specific inference instructions \cite{Su2022instructor}).  This diversity demands the creation of a verification protocol to ensure the functionality of a model prior to its inclusion in our primary DR pool.


In addition to conducting functionality checks, there is a need for infrastructural changes to scale up both computational and space resources. Current state-of-the-art models often contain billions of parameters and are very large in size, requiring the allocation of dynamic storage space. Therefore, the development of \texttt{DenseQuest} on cloud allows us to flexibly enlarge the resources in response to evolving needs. In addition, our high-level and low-level architectural design, which decouples the system into smaller components provides several benefits for future advancement. These involve improved efficiency, increased scalability, independent service development and deployment, and fault tolerance and isolation.

Finally, currently our system does not implement secure storage of the collections uploaded by users. We plan to address this by introducing privacy measures, including secure communication protocols and isolated user storage. We will also explore an option to run our algorithms locally for users with sensitive data.


\vspace{-0.15cm}
\section{Conclusion}

In this demonstration paper, we introduced \texttt{DenseQuest}, a web-based application designed to select the optimal dense retriever for private collections without requiring relevance judgments. By summarizing the architectural and infrastructural decisions, we demonstrated that our system is scalable, allowing for the parallelization of task queues to accommodate many users. Furthermore, we presented our web application, which enables real-time use of our system, and provided guiding suggestions on its effective use.
\vspace{-0.15cm}
\section*{Acknowledgment}
\label{ack}

\katya{We extend our gratitude to}
the engineer team of Thaibiogenix~Co., including Warangkhana Sukpartcharoen, Weeravat Buachoom, \\Tasanai Srisawat, and Thuchpun Apivitcholachat for their consultation and support to develop the web application for deploying  the \texttt{DenseQuest} core.

\vspace{-0.1cm}

\bibliographystyle{ACM-Reference-Format}
\bibliography{Bib/bibliography}

\end{document}